# AN EXAMINATION OF THE MULTI-PEAKED ANALYTICALLY EXTENDED FUNCTION FOR APPROXIMATION OF LIGHTNING CHANNEL-BASE CURRENTS

Karl LUNDENGÅRD[1], Milica RANČIĆ[1], Vesna JAVOR[2] and Sergei SILVESTROV[1]

**Abstract:** A multi-peaked version of the analytically extended function (AEF) intended for approximation of multi-peaked lightning current waveforms will be presented along with some of its basic properties. A general framework for estimating the parameters of the AEF using the Marquardt least-squares method (MLSM) for a waveform with an arbitrary (finite) number of peaks as well as a given charge transfer and specific energy will also be described. This framework is used to find parameters for some common single-peak wave-forms and some advantages and disadvantages of the approach will be discussed.

## INTRODUCTION

Many different types of systems, objects and equipment are susceptible to damage from lightning discharge. Lightning effects are usually analysed using lightning discharge models. Most of these models imply channel-base current functions. Various single and multi-peaked functions have been proposed in the literature, e.g. [1]-[5]. For engineering and electromagnetic models, a general function that would be able to reproduce desired waveshapes is needed, such that analytical solutions for its derivatives, integrals, and integral transformations exist. A multi-peaked channel-base current function has been proposed in [4] as a generalization of the so-called TRF (two-rise front) function from [5], which possesses such properties.

In this paper we analyse a proposed multi-peaked function, the so-called *p*-peak analytically extended function (AEF), giving explicit expressions for a number of basic properties such as derivatives and integrals.

Some discussion on how to fit a *p*-peak AEF will be done and a general framework for how paramters can be estimated using the Marquardt least-squares method is presented. Some numerical results are presented, including those for the Standard IEC62305 currents of the first negative stroke ([10]), and an example of a fast-decaying lightning current waveform. Based on presented results, corresponding conclusions are made, and further research ideas are discussed.

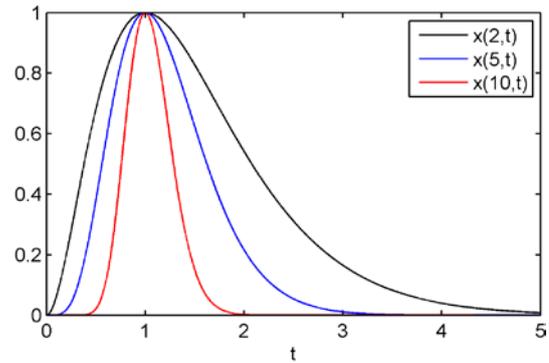

**Fig.1** – *Illustration of the power exponential function given by (1) for a few different values of β.*

## THE MULTI-PEAKED AEF

The *p*-peaked AEF is constructed using the function

$$x(\beta;t) = \left(te^{1-t}\right)^\beta, \quad 0 \leq t, \qquad (1)$$

which we will refer to as the power exponential function. The power exponential function is qualitatively similar to the desired waveforms in the sense that it has a steeply rising initial part followed by a more slowly decaying part. The steepness of both the rising and decaying part is determined by the *β*-parameter. This is illustrated in Fig. 1.

In order to get a function with multiple peaks and where the steepness of the rise between each peak as well as the slope of the decaying part is not dependent on each other, we define the analytically extended function (AEF) as a function that consist of piecewise linear combinations of the power exponential function that have been scaled and translated so that the resulting function is continuous. Given the difference in height between each pair of peaks $I_{m_1}, I_{m_2}, \ldots, I_{m_p}$, the corresponding times $t_{m_1}, \ldots, t_{m_p}$, integers $n_q > 0$, real values $\beta_{q,k}, \eta_{q,k}$, $1 \leq q \leq p+1$, $1 \leq k \leq n_q$ such that the sum over *k* of $\eta_{q,k}$ is equal to unit, the *p*-peaked AEF *i(t)* is given by Eqn. (2).

[1] Mälardalen University, Division of Applied Mathematics, UKK, Högskoleplan 1, Box 883, 721 23 Västerås, Sweden,
  e-mail: {karl.lundengard, milica.rancic, sergei.silvestrov}@mdh.se
[2] University of Niš, Faculty of Electronic Engineering, Aleksandra Medvedeva 14, 18000 Niš, Serbia, e-mail: vesna.javor@elfak.ni.ac.rs

$$i(t) = \begin{cases} \left(\sum_{k=1}^{q-1} I_{m_k}\right) + I_{m_q} \sum_{k=1}^{n_q} \eta_{q,k} x_{q,k}(t), & t_{m_{q-1}} \leq t \leq t_{m_q}, \\ & 1 \leq q \leq p, \\ \left(\sum_{k=1}^{p} I_{m_k}\right) \sum_{k=1}^{n_{p+1}} \eta_{p+1,k} x_{p+1,k}(t), & t_{m_p} \leq t, \end{cases} \quad (2)$$

with

$$x_{q,k}(t) = x\left(\beta_{q,k}^2 + 1; \frac{t - t_{m_{q-1}}}{t_{m_q} - t_{m_{q-1}}}\right), \quad 1 \leq q \leq p,$$

$$x_{p+1,k}(t) = x\left(\beta_{p+1,k}^2; \frac{t - t_{m_{q-1}}}{t_{m_q} - t_{m_{q-1}}}\right). \quad (3)$$

Introducing the vectors

$$\boldsymbol{\eta}_q = \begin{bmatrix} \eta_{q,1} & \eta_{q,2} & \cdots & \eta_{q,n_q} \end{bmatrix}^T,$$
$$\boldsymbol{x}_q(t) = \begin{bmatrix} x_{q,1}(t) & x_{q,2}(t) & \cdots & x_{q,n_q}(t) \end{bmatrix}.$$

the $p$-peaked AEF can be written in a more compact form

$$i(t) = \begin{cases} \left(\sum_{k=1}^{q-1} I_{m_k}\right) + I_{m_q} \boldsymbol{\eta}_q^T \boldsymbol{x}_q(t), & t_{m_{q-1}} \leq t \leq t_{m_q}, \\ & 1 \leq q \leq p, \\ \left(\sum_{k=1}^{p} I_{m_k}\right) \boldsymbol{\eta}_{p+1}^T \boldsymbol{x}_{p+1}(t), & t_{m_p} \leq t. \end{cases} \quad (4)$$

## THE DERIVATIVE AND INTEGRAL OF THE ANALYTICALLY EXTENDED FUNCTION

Since the $p$-peaked AEF is constructed from elementary functions, it is very straightforward to find the derivative with standard methods. The resulting expression is

$$\frac{di}{dt} = \begin{cases} I_{m_q} \frac{t_{m_q} - t}{t - t_{m_{q-1}}} \frac{x_q(t)}{\Delta t_{m_q}} \boldsymbol{\eta}_q^T \boldsymbol{B}_q \boldsymbol{x}_q(t), & t_{m_{q-1}} \leq t \leq t_{m_q}, \\ & 1 \leq q \leq p, \\ I_{m_q} \frac{t_{m_q} - t}{t_{m_q}} \frac{x_q(t)}{\Delta t_{m_q}} \boldsymbol{\eta}_q^T \boldsymbol{B}_q \boldsymbol{x}_q(t), & t_{m_p} \leq t, \\ & q = p+1, \end{cases} \quad (5)$$

where $\boldsymbol{B}_q$ are diagonal matrices

$$\boldsymbol{B}_q = diag(\beta_{q,1}^2 + 1, \beta_{q,2}^2 + 1, \ldots, \beta_{q,n_q}^2 + 1), \quad 1 \leq q \leq p,$$

$$\boldsymbol{B}_{p+1} = diag(\beta_{p+1,1}^2, \beta_{p+1,2}^2, \ldots, \beta_{p+1,n_{p+1}}^2).$$

Based on this expression, it is easy to see that the first derivative is continuous since it will be zero at each $t_{m_q}$.

The integral of the AEF is also relatively easy to find, since the integral of the power exponential function can be written using the lower incomplete gamma function, which is a well-known special function, see for example [7]. More specifically

$$\int_{t_0}^{t_1} x(\beta;t)\, dt = \frac{e^{\beta-1}}{\beta^\beta}\left(\gamma(\beta+1;\beta t_1) - \gamma(\beta+1;\beta t_0)\right), \quad (6)$$

where $\gamma(\beta,t) = \int_0^t \tau^{\beta-1} e^{-\tau} d\tau$ is the incomplete gamma function.

Using Eqn. (6) and Eqn. (2) together with some standard results for integration gives that for $0 \leq t_a \leq t_b \leq t_{m_p}$, $t_{m_{a-1}} \leq t_a \leq t_{m_a}$, $t_{m_{b-1}} \leq t_b \leq t_{m_b}$, the integral of the rising part of the AEF is

$$\int_{t_a}^{t_b} i(t) = (t_{m_a} - t_a)\left(\sum_{k=1}^{a-1} I_{m_k}\right) + I_{m_a} \sum_{k=1}^{n_a} \eta_{a,k} g_a(t_a, t_{m_a}) +$$
$$+ \sum_{q=1}^{b-1}\left(\Delta t_{m_q}\left(\sum_{k=1}^{q-1} I_{m_k}\right) + I_{m_q} \sum_{k=1}^{n_q} \eta_{a,k} \hat{g}\left(\beta_{q,k}^2 + 1\right)\right) + \quad (7)$$
$$+ (t_{m_a} - t_a)\left(\sum_{k=1}^{a-1} I_{m_k}\right) + I_{m_b} \sum_{k=1}^{n_q} \eta_{b,k} g_b(t_{m_b}, t_b),$$

where $g_q(t_1, t_2)$ is the integral of the power exponential function $x(\beta_q;t)$ given by Eqn. (6) and

$$\hat{g}(\beta) = \frac{e^{\beta-1}}{\beta^\beta}\gamma(\beta+1,\beta).$$

Similarly for the decaying part, $t_{m_p} \leq t_0 \leq t_1 < \infty$, the integration formula is

$$\int_{t_0}^{t_1} i(t)\, dt = \left(\sum_{k=1}^{p} I_{m_k}\right) \sum_{k=1}^{n_{p+1}} \eta_{p+1,k} g_{p+1}(t_1, t_0). \quad (8)$$

## FITTING THE AEF TO DATA USING THE MARQUARDT LEAST SQUARE METHOD

Here we will attempt to fit the AEF to some different current waveshapes. We will use the Marquardt least square method (MLSM) to estimate the $\beta$-parameters, and from these calculate the corresponding $\eta$–paramaters.

We will not give a detailed description of MLSM, instead we point to [9] for a description of how to apply it in a similar situation. Here we will only supply the parts of the method specific to the use of the AEF.

The MLSM uses a Jacobian matrix that contains the partial derivatives of the residuals, here we denote this matrix with $\boldsymbol{J}$. Suppose that we want to find the least square fit of the AEF to a set of data points. Then the fitting can be done separately between each peak (and after the final peak). The $\boldsymbol{J}$ matrix in this case is

$$\boldsymbol{J} = \begin{bmatrix} p_{q,1}(t_{q,1}) & p_{q,2}(t_{q,1}) & \cdots & p_{q,n_q}(t_{q,1}) \\ p_{q,1}(t_{q,2}) & p_{q,2}(t_{q,2}) & \cdots & p_{q,n_q}(t_{q,2}) \\ \vdots & \vdots & \ddots & \vdots \\ p_{q,1}(t_{q,k_q}) & p_{q,2}(t_{q,k_q}) & \cdots & p_{q,n_q}(t_{q,k_q}) \end{bmatrix}, \quad (9)$$

where $k_q$ is the number of data points between the $q$th and $(q$-$1)$th peak, $t_{q,r}$ are the corresponding times for these data points, and $p_{q,r}(t_{q,s}) = \left.\frac{\partial i}{\partial \beta_{q,r}}\right|_{t=t_{q,r}}$ more explicitly given by

$$p_{q,r}(t_{q,s}) = 2I_{m_q}\eta_{q,k}\beta_{q,k}h_q(t)x(\beta_{q,k}^2+1), 1 \le q \le p,$$

$$p_{p+1,r}(t_{q,s}) = I_{m_{p+1}}\eta_{a+1,k}\beta_{p+1,k}h_q(t)x(\beta_{p+1,k}^2).$$

where

$$h_q(t) = \ln\left(\frac{t-t_{m_{q-1}}}{\Delta t_{m_q}}\right) + \frac{t-t_{m_{q-1}}}{\Delta t_{m_q}} + 1, 1 \le q \le p,$$

$$h_{p+1}(t) = \ln\left(\frac{t}{t_{m_{p+1}}}\right) + \frac{t}{t_{m_{p+1}}} + 1.$$

The MLSM is an interative method, and in order to find a new set of β-parameters in each iteration we also need to find the η–parameters. This is done by using the regular least square method since for fixed β the AEF is linear in η.

Often we also wish to take the charge transfer, $Q_0$, and specific energy, $W_0$, of the lightning discharge into account. These two quantities are given by

$$Q_0 = \int_0^\infty i(t)\,dt, \qquad (10)$$

$$W_0 = \int_0^\infty i(t)^2\,dt. \qquad (11)$$

Using Eqns. (7) and (8) the corresponding quantities for the AEF are found to be

$$Q = \sum_{q=1}^{p}\left(\Delta t_{m_q}\left(\sum_{k=1}^{q-1}I_{m_k}\right) + I_{m_q}\sum_{k=1}^{n_q}\eta_{q,k}\,\hat{g}(\beta_{q,k}^2+1)\right) + \\ +\left(\sum_{k=1}^{p}I_{m_k}\right)\sum_{k=1}^{n_{p+1}}\eta_{p+1,k}\,\tilde{g}(\beta_{p+1,k}^2), \qquad (12)$$

$$W = \sum_{q=1}^{p}W_q + \left(\sum_{k=1}^{p}I_{m_k}\right)W_{p+1}, \qquad (13)$$

where

$$\tilde{g}(\beta) = \frac{e^{\beta-1}}{\beta^\beta}(\Gamma(\beta+1) - \gamma(\beta+1,\beta)),$$

$$W_q = \Delta t_{m_q}^2\left(\sum_{k=1}^{q-1}I_{m_k}\right)^2 + I_{m_q}^2\sum_{k=1}^{n_q}\eta_{q,k}\,\hat{g}(\beta_{q,k}^2+1) + \\ + I_{m_q}\Delta t_{m_q}\left(\sum_{k=1}^{q-1}I_{m_k}\right)\sum_{k=1}^{n_q}\eta_{q,k}\,\hat{g}(\beta_{q,k}^2+1) + \\ + 2I_{m_q}^2\sum_{r=1}^{n_q-1}\sum_{s=r+1}^{n_q}\eta_{q,r}\eta_{q,s}\hat{g}(\beta_{q,r}^2+\beta_{q,s}^2+2), 1 \le q \le p,$$

$$W_{p+1} = \sum_{k=1}^{n_q}\eta_{q,k}^2\,\tilde{g}(\beta_{p+1,k}^2) + \\ + 2I_{m_q}^2\sum_{r=1}^{n_q-1}\sum_{s=r+1}^{n_q}\eta_{p+1,r}\eta_{p+1,s}\tilde{g}(\beta_{p+1,r}^2+\beta_{p+1,s}^2).$$

Since the charge transfer and specific energy place conditions on the entire function it is no longer possible to fit each interval separately. Instead the entire function is fitted at once using the **J** matrix

$$J = \begin{bmatrix} J_1 & 0 & \cdots & 0 \\ 0 & J_2 & \cdots & 0 \\ \vdots & \vdots & \ddots & \vdots \\ 0 & 0 & \cdots & J_{p+1} \\ \hat{Q}_1 & \hat{Q}_2 & \cdots & \hat{Q}_{p+1} \\ \hat{W}_1 & \hat{W}_2 & \cdots & \hat{W}_{p+1} \end{bmatrix},$$

where $J_q$, $1 \le q \le p+1$, is the matrix for the corresponding interval and $\hat{Q}_q$ and $\hat{W}_q$, $1 \le q \le p+1$, are matrices with the partial derivatives of the $Q$ and $W$ with respect to the β-parameters.

The expressions for the partial derivatives are somewhat complicated and have here been omitted to save space. Full expressions can be found in [12].

Note that the residuals for the charge transfer and specific energy can be on a very different scale from the residuals of the current values. Therefore some sort of weighted least square fitting might be more appropriate in many situations.

## RESULTS AND DISCUSSION

Here we will show some results of the fitting for 1-peak and 2-peak AEF functions to data either given by Heidler functions (Fig. 2, Fig. 3 and Fig. 4) or experimental data (Fig. 5).

In Fig. 2 and Fig. 3 monotonicity of the functions have been guaranteed by forcing all η–parameters to be positive. This was not necessary for Fig. 4 and Fig. 5.

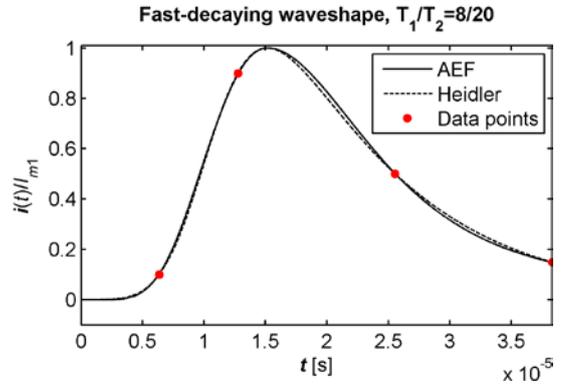

**Fig.2** – *Illustration of the 1-peak AEF fitted to four data-points generated by the Heidler function, note the good fit.*

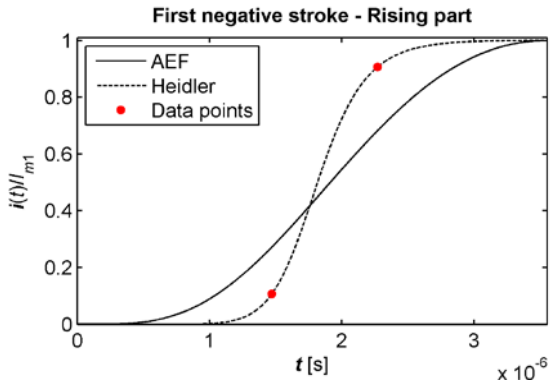

**Fig.3** – *Illustration of the 1-peak AEF fitted to two data-points generated by the Heidler function, note the poor fit.*

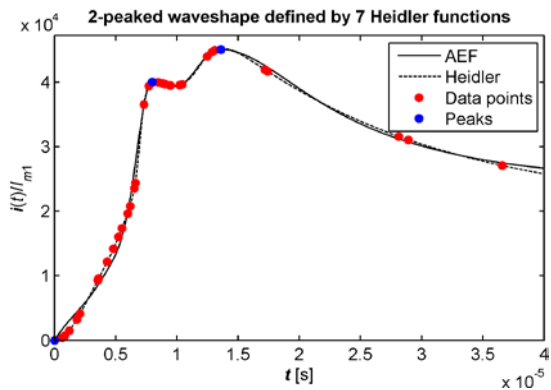

**Fig.4** – *The 2–peak AEF with 3 terms in each interval fitted to random data from a waveshape given by 7 Heidler functions.*

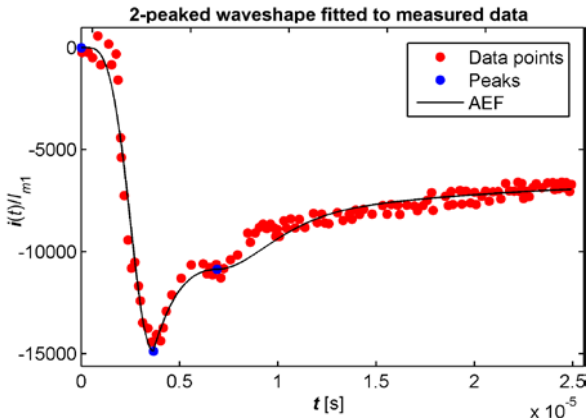

**Fig.5** – *The 2–peak AEF with 2, 1 and 4 terms in each interval fitted to measurement data from [11].*

In Fig. 2 the fitting for fast-decaying waveshape is shown, and in Fig. 3 the fitting for rising part of the first negative stroke. The reason that the fit in Fig. 3 is so poor is due to the fact that increasing the steepnees of the AEF also moves the transition forward in time. Thus we would need either to move the time for the peak or complement the AEF with some function that can have a steep rise in the middle of the interval.

In Fig. 4 the quality of the fit varies between the interval. In the first interval the fit is not very good but in the second interval the AEF approximates the waveshape well. We can even use the AEF to approximate the peak in the middle of the interval without need to specify its position. Since the data points were chosen randomly for this fitting and the results varied considerably depending on chosen points some strategy for choosing points should be devised.

In Fig. 5 the AEF fits the data well for the most part and here further experimentation with the number of terms in each interval could be fruitful.

It should be noted that in all cases MLSM showed some tendencies to find local minima in the objective function instead of the global and therefore other methods of fitting should be explored, such as genetic algorithms or more specialised methods, for instance methods used for $L$-splines that the AEF is conceptually similar to.